\documentclass[3p,times,twocolumn]{elsarticle}
 \biboptions{comma,sort&compress}
 
\usepackage{graphicx}
\usepackage{here}
\usepackage{ecrc}

\volume{00}
\firstpage{1}
\journalname{Nuclear and Particle Physics Proceedings}
\runauth{}
\jid{nppp}
\jnltitlelogo{Nuclear and Particle Physics Proceedings}
\usepackage{amssymb}
\usepackage[figuresright]{rotating}

\begin{document}

\begin{frontmatter}

%%
%%%%%%%%%%%%%%%%%%%%%%%%%%%%%%%%%%%%%%%%%%%%%%%%%
\title{New developments in model-independent Partial-Wave Analysis${}^*$}
 % \corref{cor0}}
 \cortext[cor0]{Talk given at 23rd International Conference in Quantum Chromodynamics (QCD 20,  35th anniversary),  27 october - 30 october 2020, Montpellier - FR by F.~Krinner}
 \author[label1]{F.~M.~Krinner}
%  \cortext[cor0]{FAPESP CNPq-Brasil PhD student fellow.}
   \fntext[fn1]{Speaker, Corresponding author.}
\ead{fkrinner@mpp.mpg.de}
 \author[label1,label2]{S.~Paul}
\address[label1]{Max Planck Institut f\"ur Physik, F\"ohringer Ring 6, 80805 M\"unchen, Germany}
\address[label2]{Technische Universit\"at M\"unchen, James-Franck-Straße 1, 85748 Garching, Germany}
\pagestyle{myheadings}
\markright{ }
\begin{abstract}
Partial-wave analyses (PWA) are an essential tool for studying 
resonance structures in decays with hadronic multi-body final 
states. For several years, more model-independent approaches to such 
analyses have been used for various decay final states. However, 
up to now, these methods have mostly been applied to sub-sets of 
partial waves, also called freed waves. In this article, we 
explore possibilities and limitations of extended model-independent 
approaches. We systematically apply various different fit models 
to the analysis of pseudo data sets, to study both the impact of 
the mathematical description used for the freed waves and the 
choice of simultaneously freed waves. We can show that suitable 
methods exist, which lift restrictions to only sub-sets of freed 
partial waves and demonstrate hidden caveats present in previous 
works.
%Partial-wave analyses (PWA) are an essential tool  
%for studying resonance structures in decays with hadronic 
%multibody final states. For several years, model-independent 
%approaches to this type of analysis have been applied to 
%various decays. However, up to now, these methods have mostly
%been used on sub-sets of partial waves.     
%
%In this article, we explore possibilities and limitations
%of extended model-independent approaches. We systematically
%apply various different fit models on a pseudo data
%set, and study the impact of the binning  of the invariant 
%mass of a particle subsystem of the final state, the choice 
%of the functional description in the subsystem and of the 
%choice waves subjected to the model independent fits.
\end{abstract}
% \begin{document}

\end{frontmatter}

\section{Introduction}
Amplitude analyses of hadronic multi-body final states 
prove to be a valuable tool to disentangle the 
excitation spectrum of QCD and to determine strong phases 
in weak decays. Small contributions to the decay amplitude 
can interfere with dominant ones and thereby become visible 
and accessible through amplitude analysis. However, 
Partial-Wave Analyses (PWA) of multi-body final-states, 
which constitute the most common tool for amplitude
analyses, require amplitude modelling. The assumptions 
underlying the models can introduce a model bias and
possibly obscure 
small signal components. In this article we propose 
improvements to the so-called ``model-independent PWA'' 
approach often used to alleviate the necessity for such 
model assumptions; other names for this method include 
``freed-isobar PWA'' or ``quasi-model-independent PWA''. 
In particular, we explore the suppression of ambiguities, 
arising when extending the model-independent approach to 
multiple parts of the amplitude model.

\section{Model-independent partial-wave analysis}
\label{sec:mipwa}
To understand the complex-valued amplitudes governing a 
particular meson decay, we use the framework of partial-wave 
analysis. Here, the measured intensity distribution 
$\mathcal{I}\left(\vec\theta\,\right)$---which 
for three-body decays often is 
depicted in a Dalitz plot---depends on the kinematic 
variables $\vec\theta$ and is described as the modulus 
square of the full modeled amplitude:
\begin{equation}\label{eq:intens}
\mathcal{I}\left(\vec\theta\,\right) = \left|\sum_{w\in\mathrm{waves}}\,\mathcal{T}_w\,\mathcal{A}_w\left(\vec\theta\,\right)\right|^2
,\end{equation}
where the full amplitude is decomposed into a sum over a 
set of partial waves $w$. The complex-valued transition 
amplitudes $\mathcal{T}_w$ encode the strength and 
relative phases of the individual partial waves $w$; they 
are the free parameters of a PWA. In contrast, the complex-valued 
decay amplitudes $\mathcal{A}_w\left(\vec\theta\,\right)$ usually 
do not contain any free parameters, but encode the dependence of 
the individual partial waves on $\vec\theta$. They have to be 
fully known beforehand, which requires modelling and prior 
knowledge.

For the description of the decay amplitudes, we use the isobar 
model, which describes a multi-body decay process as a sequence of 
subsequent two-particle decays, which includes the appearance of  
resonant sub-systems, the isobars. In this article, we focus on 
a three-body final state, where the process is described by 
two two-body decays introducing an additional two-particle 
sub-system $\xi$. Within this isobar model, we express the 
decay amplitudes $\mathcal{A}_w\left(\vec\theta\,\right)$ as:
\begin{equation}\label{eq:pwa}
\mathcal{A}_w\left(\vec\theta\,\right) = \psi_w\left(\vec\theta\,\right) \Delta_w\left(m_{\xi}^2\right)
.\end{equation}
The angular amplitudes $\psi_w\left(\vec\theta\,\right)$ describe 
the dependence of the partial waves on the decay angles, which is 
given by first principles and fully determined by the spin and 
angular momentum quantum numbers of the corresponding partial 
wave $w$. The dynamic amplitude $\Delta_w\left(m_{\xi}^2\right)$ 
is a complex-valued function of only the invariant mass of the 
isobar system and describes its resonance content. They are most 
commonly parameterized through Breit-Wigner functions describing 
isolated resonances appearing far from thresholds.
%The most common choice 
%to model dynamic amplitudes is the Breit-Wigner shape, which describes 
%an isolated resonance far from thresholds. 
However, the appearance of more than one such resonance with the 
same $J^{PC}$ quantum numbers or interactions between $\xi$ and 
the third spectator particle require more advanced modelling. The actual model---and 
its model parameters---are not 
given by first principles. At any rate, the dynamic amplitude is a 
necessary input for a partial wave model.

To minimize the number of assumptions for the dynamic amplitudes,
model-independent PWA approaches are often used. Here, 
the mass distribution of the isobar $m^2_\xi$ is discretized into 
bins $b$ and the complex valued dynamic amplitude is 
replaced by a set of basis functions~$\Lambda^b\left(m^2_{\xi}\right)$,
defined within a mass bin $b$:
\begin{equation}\label{eq:mipwa}
\Delta_w\left(m_{\xi}^2\right) = \sum_{b\in\mathrm{bins}} \mathcal{C}_w^b \Lambda_w^b\left(m^2_{\xi}; m^2_{b-1}, m^2_b, m^2_{b+1}\right)
.\end{equation}
The full dynamic amplitude is obtained by summing of all bins $b$ in 
the invariant mass-squared $m_{\xi}^2$ of the isobar, that 
contiguously span the full kinematically allowed 
range. In the past, piecewise constant functions were most commonly 
used as function basis 
\cite{e791,compass,lhcb}. However, in our studies we found spike 
functions, 
$\Lambda^b\left(m^2_{\xi}; m^2_\mathrm{low},m^2_\mathrm{center}, m^2_\mathrm{up}\right)$ 
growing linearly from 0 to 1 as $m_\xi^2$ goes from 
$m^2_\mathrm{low}$ to $m^2_\mathrm{center}$ and within a mass bin 
then returning to 0 as $m_\xi^2$ reaches $m_\mathrm{up}^2$ to 
result in a more favorable convergence behavior. The spike 
functions vanish outside the range 
$]m^2_\mathrm{low},m^2_\mathrm{up}[$.\footnote{For the first and the 
last bin, $m^2_\mathrm{low}$ and $m_\mathrm{up}^2$ lie outside the 
kinematically allowed range and are therefore neglected.}

The complex-valued coefficients $\mathcal{C}_w^b$ in Eq.~\ref{eq:mipwa} 
encode the value of the dynamic amplitude in a given $m_{\xi}^2$ bin $b$. 
Note, that the replacement given in Eq.~(\ref{eq:mipwa}) does not alter 
the over-all structure of Eq.~(\ref{eq:intens})\footnote{The free 
parameters in such a fit are the products $\mathcal{T}_w\mathcal{C}_w^b$, 
where the $\mathcal{T}_w$ are a global complex-valued factor for the 
dynamic isobar amplitude, which acts as a global phase- and normalization 
factor.}. Covering the kinematically allowed $m^2_{\xi}$ range with 
overlapping spike functions results in an approximation to the dynamic 
amplitude, which is linear in $m_\xi^2$.

The use of higher order polynomials as basis functions instead of a linear 
interpolation across the isobar mass bins, ensures also derivatives of 
the dynamic amplitudes to be continuous at the bin borders. Such 
polynomials of degree $n$ have to be non-vanishing in $n+1$ neighboring 
bins, which results in $n+1$ non-vanishing basis functions 
at any given value of $m_{\xi}^2$ and thus to large overlaps between the 
functions. Such large overlaps tend to produce artifacts, large 
correlations among the fit parameters in the analysis, and a worse convergence 
towards the input model. Function bases, that don't rely on a fixed binning 
in $m_{\xi}^2$, like e.g. the Fourier basis, or Chebyshev polynomials 
suffer from similar problems.

\section{The pseudo data sets}
\label{sec:dataSet}

To study the viability and limitations of the model-independent PWA 
approach, we chose the most simple case in form of a three-body decay 
of a spin 0 initial state into three spin 0 final-state particles, 
which can be depicted in a Dalitz plot. 
However, we stress that the methods discussed are viable for a large 
variety of PWA of three-body final states. We will exemplarily consider the 
decay
\begin{equation}
\mathrm{D}^+ \to \mathrm{K}^- + \pi^+ + \pi^+
\end{equation}
for which we generated two pseudo data sets using Monte Carlo methods. 
The decay is modeled with three partial 
waves with the spin of the $[\mathrm{K}^-\pi^+]$ isobar ranging from 
zero to two. Since all initial and final state particles are spinless, 
the spin of the isobar already fixes all quantum numbers of a partial 
wave and the set of kinematic variables is given by:
\begin{equation}
\vec\theta = \{m_{\mathrm{K}\pi_1}^2, m_{\mathrm{K}\pi_2}^2 \}
.\end{equation}
For a data set (A) of a million events, the S-, P-, and D-wave we 
used the following $[\mathrm{K}^-\pi^+]$ resonances to describe the 
isobars:
\begin{equation}
\mathrm{K}^*_0(700)\ ;\ \quad\mathrm{K}^*(892)\ ;\ \quad\mathrm{K}_2^*(1430)
.\end{equation}
The appearance of two identical $\pi^+$ in the final state, requires 
the amplitude in Eq.~(\ref{eq:pwa}) to be symmetrized under the 
exchange of the two $\pi^+$ to fulfill Bose symmetry. We do not use 
doubly charged isobars. 
%Fig.~\ref{fig:dalitz} shows the resulting 
%Dalitz plot, where the dominating $\mathrm{K}^*(892)$ can easily be 
%identified.
%
%\begin{figure}[htb]
%{\includegraphics[width=.85\columnwidth]{./plots/dalitz}}
%\caption{Dalitz plot for the pseudo data set with the excited 
%isobar resonances $\mathrm{K}_0^*(1430)$ and $\mathrm{K}^*(1410)$ 
%included.}
%\label{fig:dalitz}
%\end{figure}

The validity of the model-independent PWA method, which aims at 
reconciling the isobar structure, is studied through a 
second pseudo data set (B) of the same size now also including a 
second resonance in both the S- and P- wave, namely the 
$\mathrm{K}_0^*(1430)$ and $\mathrm{K}^*(1410)$ isobars. For all 
appearing resonances we use Breit-Wigner parameterizations with 
resonance parameters taken from Ref.~\cite{pdg}. 

\section{Model-independent PWA results}
We now perform an analysis on the two pseudo data-sets 
generated in Sec.~\ref{sec:dataSet}.

\subsection{Fit model}
\label{sec:fitModel}
We model the dynamic amplitudes of the S- and P-wave in Eq.~(\ref{eq:intens})
by spike functions as given in Eq.~(\ref{eq:mipwa}). However, instead of 
an equidistant binning in the isobar mass we optimize the 
bin borders $\{m^2_b\}$ for a given number of bins to such that 
Eq.~(\ref{eq:mipwa}) minimizes the integral of 
the difference of fixed shape amplitudes $\Delta^{\mathrm{BW}}_w(m^2_\xi)$ 
and the model independent approximation:
\begin{equation}\label{eq:integral}
\!\!\int\!\!\!\mathrm{d}m^2 \left|\Delta^{\mathrm{BW}}_w\!(m^2)\!-\!\! \sum_b\!\mathcal{C}_w^b \Lambda_w^b\!\left(m^2;m^2_{b-1},m^2_{b},m^2_{b+1}\right) \right|^2\!\!
.\end{equation}
Here, the coefficients $\mathcal{C}_w^b$ are adjusted simultaneously 
with $\{m_b^2\}$ to guarantee the best approximation for any binning by.
We can determine the optimal number of mass bins $n_\mathrm{Opt}$ 
for a given data set by Monte-Carlo studies. We optimize towards 
resolving resonant features in the spectral distribution of $m^2_\xi$ 
while avoiding overfitting caused by fluctuations in the scarcely 
populated part of the mass spectrum. Using $10^6$ pseudo events, 
$n_\mathrm{opt} = 20$ for our decay models. This number was 
determined by comparing likelihood values for the input model and 
the model independent fit. It depends on the size of the data set.

\begin{figure}[htb]
{\includegraphics[width=.85\columnwidth]{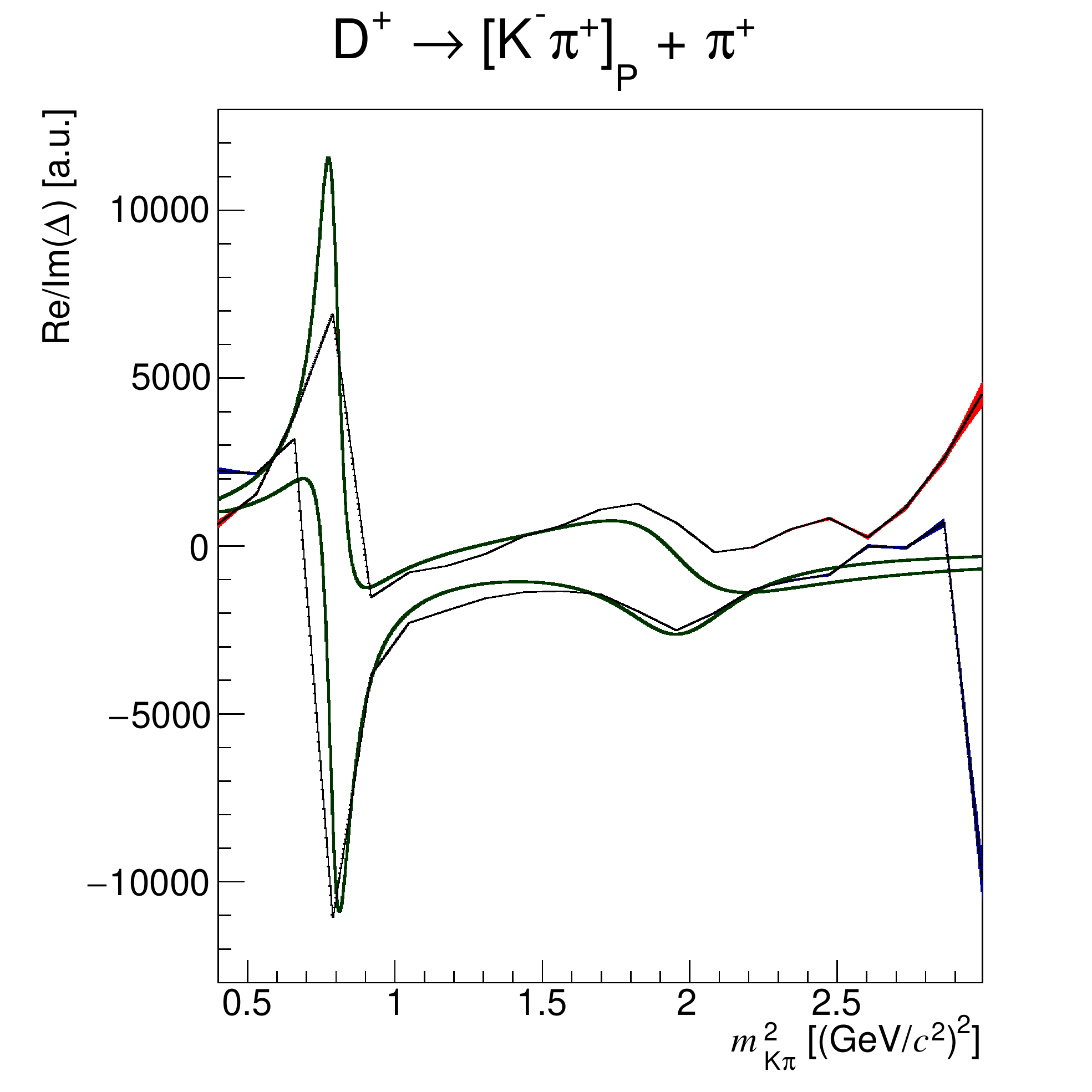}}
\caption{Model independent PWA fit result for the P-wave with equidistant  
bin borders for data set B. The red and blue bands show the real and 
imaginary parts as extracted by the model-independent fits. The green 
lines show the input shape used for the  pseudo data. The width of the 
band give the statistical uncertainty of the fit.}
\label{fig:nonopt}
\end{figure}

We now compare the results from the model-independent fit with the input 
shape over the full range of $m_\xi^2$ for both, the equidistant binning 
(see Fig.~\ref{fig:nonopt}) and the optimized binning (see Fig.~\ref{fig:excitedFit}).

\subsection{Base fit}
\label{sec:baseFit}

\begin{figure}[htb]
%{\includegraphics[width=\columnwidth]{./Swave_exc}}\\
{\includegraphics[width=.85\columnwidth]{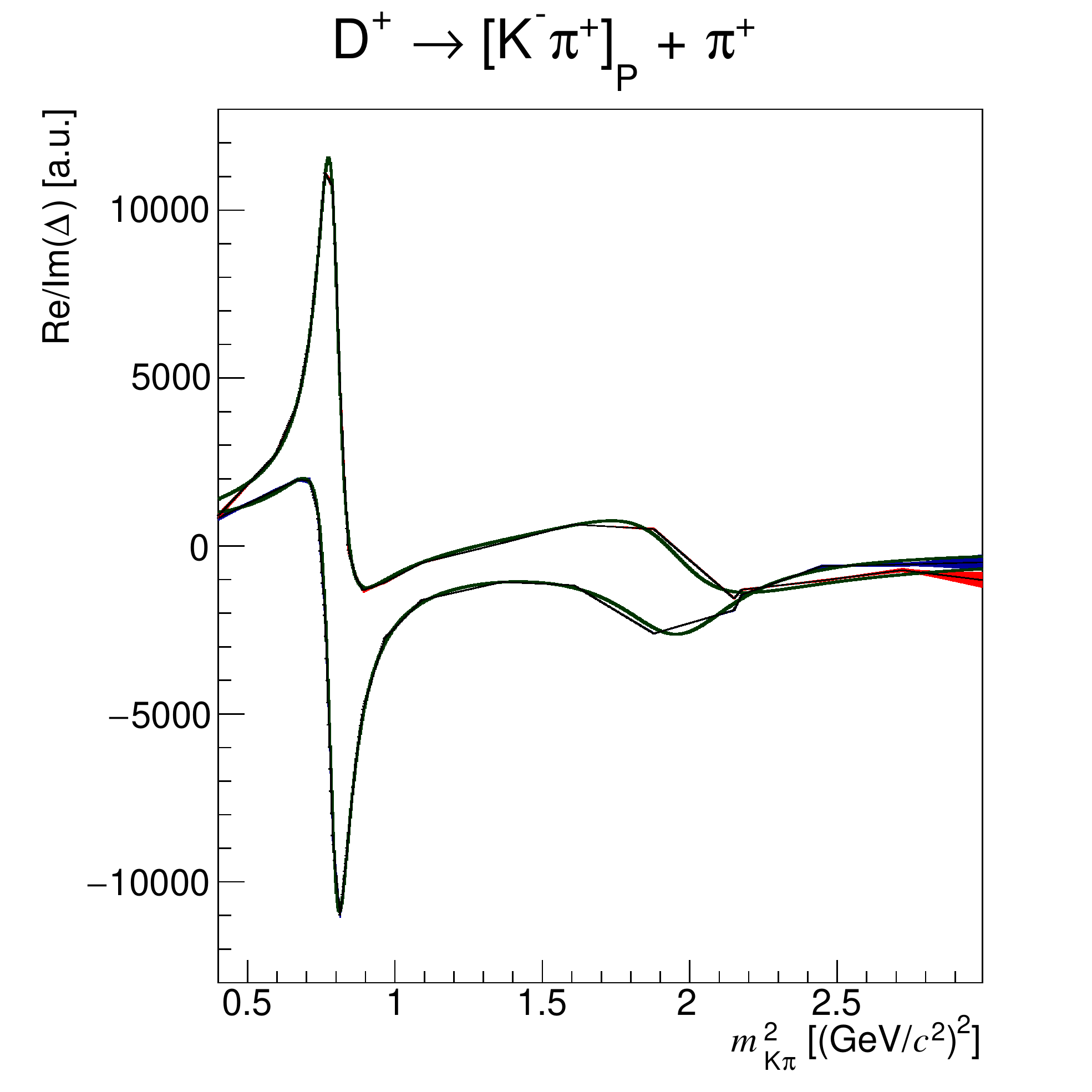}}
\caption{Result from model-independent PWA fits for the P-wave
 using pseudo data set B including several excited mesons. The 
color code is the same as in Fig.~\ref{fig:nonopt}.
}
\label{fig:excitedFit}
\end{figure}

We performed partial-wave analyses with the model independent 
approach for the S- or P-wave separately using the above 
mentioned technique with optimized binning in the isobar mass,
and apply this to the pseudo data set A, which only contains 
the ground state resonances. We obtain good agreement between 
the input model and the fit result. In order to exclude this 
agreement to be related only to the new optimized binning method, 
we applied this fit using the identical mass binning to pseudo 
data set B based on the extended model including the 
$\mathrm{K}_0^*(1430)$ and $\mathrm{K}^*(1410)$ resonances. 
Again, we find a good agreement between model input and fit 
results, as seen in Fig.~\ref{fig:excitedFit} for the P-wave, 
%Especially the $\mathrm{K}^*(1410)$ mass region is reproduced
%well, even though it was not considered when optimizing the 
%binning.
although the binning was optimized for data set A. 
This is caused by the narrow $K^*(892)$ to be well accommodated 
by a more narrow binning, while the coarser resolution away 
from this resonance suffices to pick up the significantly broader 
excited resonances.

\subsection{Cross talk}

\begin{figure}[htb]
%{\includegraphics[width=\columnwidth]{./plots/Swave_mis}}\\
{\includegraphics[width=.85\columnwidth]{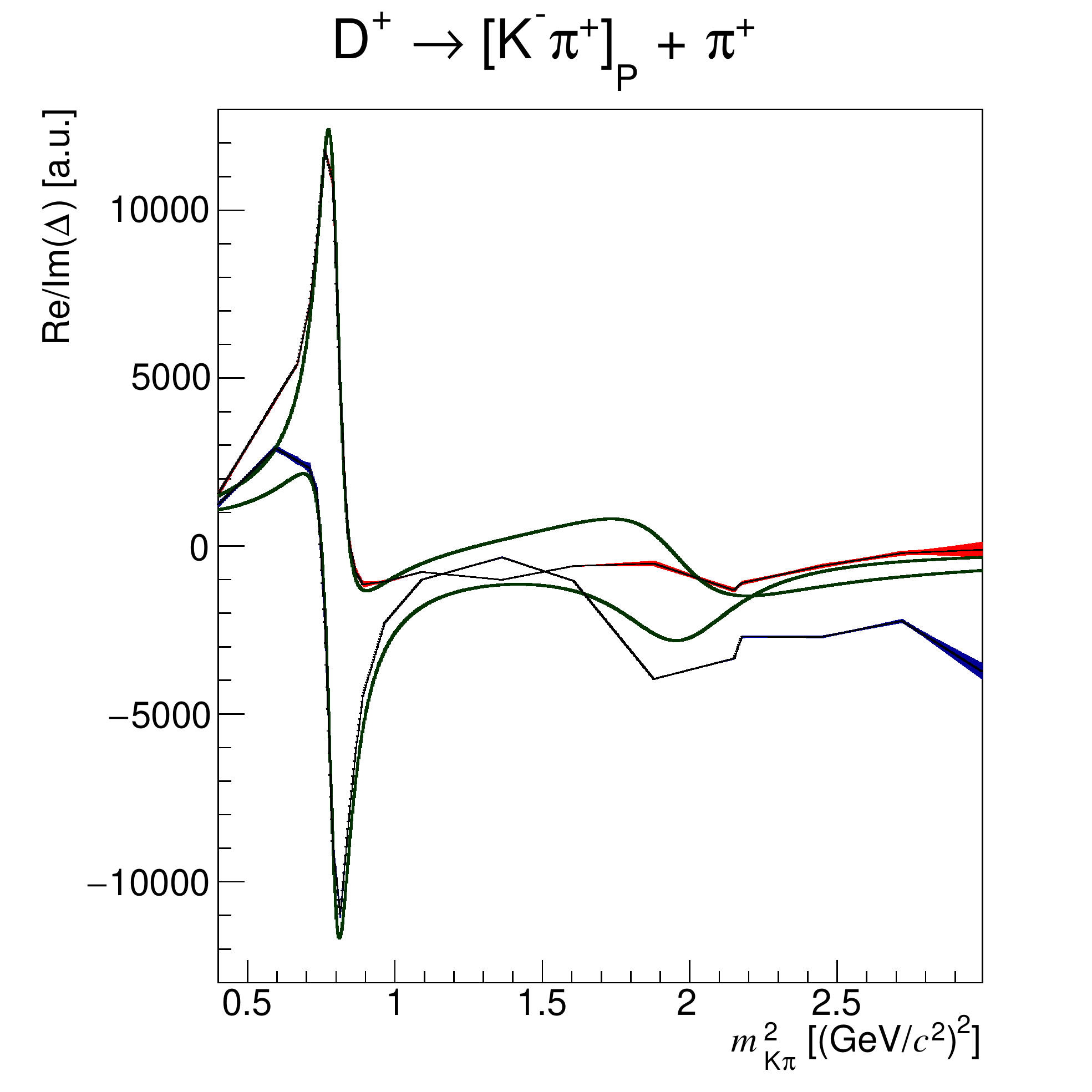}}
\caption{Result from model-independent PWA fits for the
 P-wave on a pseudo data set B including an excited 
S-wave resonance, which in turn is not included in the fit model. 
The color code is the same as in Fig.~\ref{fig:nonopt}.}
\label{fig:wrongModelFit}
\end{figure}

For the studies in Sec.~\ref{sec:baseFit} we used the input 
model amplitudes to describe those partial waves not subjected 
to the model independent formulation. It is now interesting to 
probe the method for cross talk between waves; namely we use an 
incomplete fixed model description for the S-wave and investigate 
the P-wave in a model-independent way. The results of this fit 
are shown in Fig.~\ref{fig:wrongModelFit}. Using an incomplete 
model description for the fixed S-wave impacts the results for 
the model independent P-wave, which does not reproduce the input 
data anymore.

The large freedom present in the model-independent wave allows to 
accommodate part of the difference between the input model and the
fit model for the S-wave thereby distorting the model-independent 
P-wave. It thus generates cross talk. For most meson decay analyses, 
the shapes of the amplitudes are not know a priori. Thus, they must 
be extracted from data, which in turn only gives consistent results, 
if this determination is done simultaneously in all waves, not in a 
partial or iterative way as often performed before.

However, such simultaneous approaches often generate unphysical results 
and large uncertainties due to exact mathematical cancellations of 
different amplitudes or parts of them, as can be seen for example in 
Ref.~\cite{zeroModes}. However, when correctly identified, these 
cancellations can be removed while still avoiding potential leakage 
effects from improper parameterizations of fixed waves in the model.

\begin{figure}[htb]
%{\includegraphics[width=.85\columnwidth]{./plots/Swave_both}}\\
{\includegraphics[width=.85\columnwidth]{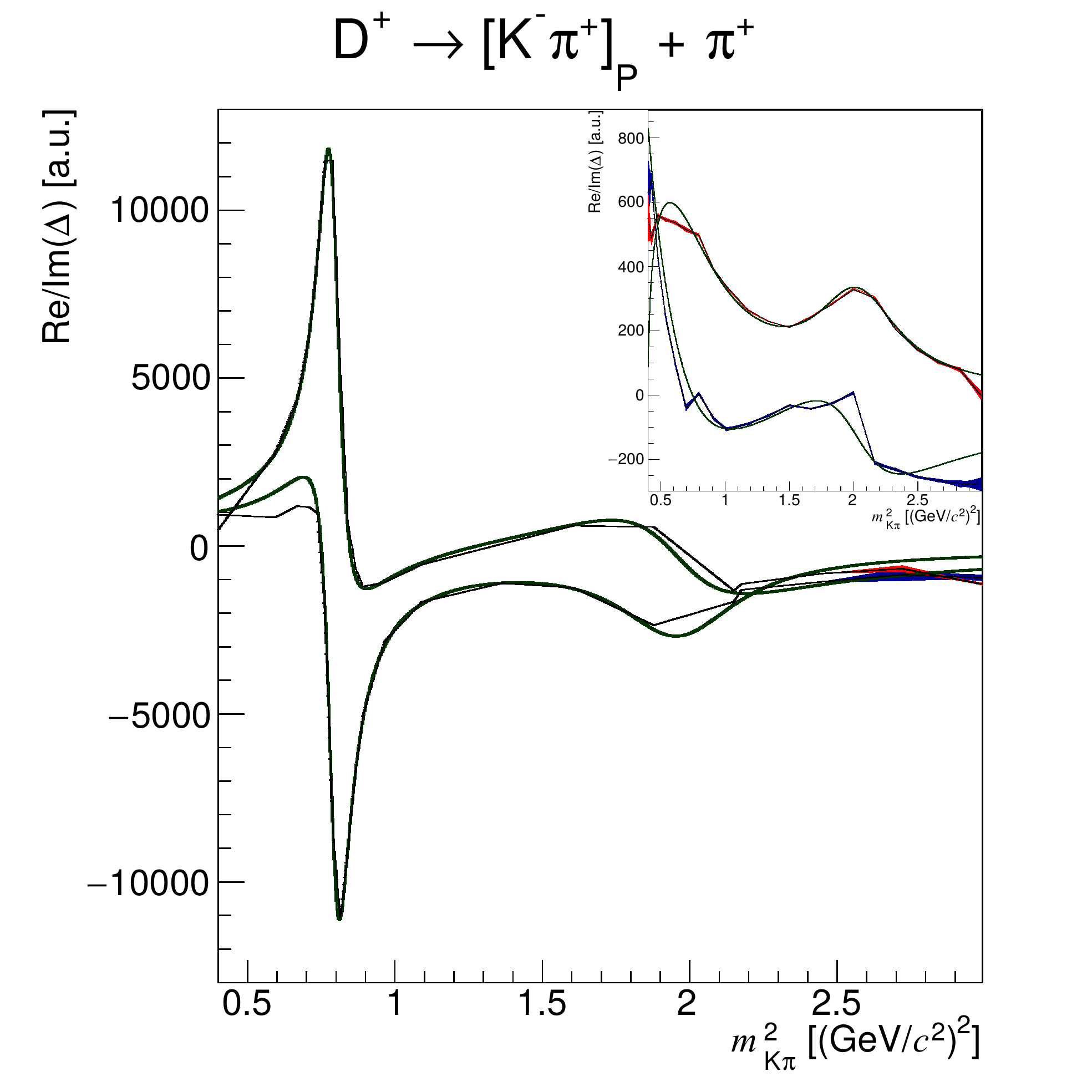}}
\caption{Result for the P-wave from a PWA fit simultaneously employing 
model-independent S- and P-waves. The color code is the same as in 
Fig.~\ref{fig:nonopt}. The corresponding result for the S-wave is
shown as inlay.}
\label{fig:simul}
\end{figure}

The origin of these cancellations are the particular angular 
amplitudes
$\psi^{L}_{12}\left(\vec\theta\,\right)$, 
for an isobar formed by particles 1 and 2. For the 
S- and P-wave they are given by:
\begin{equation}
\psi_{12}^\mathrm{S}\left(\vec\theta\,\right) \propto 1
\end{equation}
and
\begin{equation} \psi_{12}^\mathrm{P}\left(\vec\theta\,\right) \propto m^2_{23} - m^2_{13} - \left(m_{123}^2 - m_{3}^2\right) \frac{m_{1}^2 - m_{2}^2}{m_{12}^2}
.\end{equation}
Due to Bose symmetrization, the total contribution of the S- and 
P-wave to the amplitude 
$\mathcal{A}_{\mathrm{S} + \mathrm{P}}\left(\vec\theta\,\right)$
is given by:
\begin{eqnarray}
\mathcal{A}_{\mathrm{S} + \mathrm{P}}\left(\vec\theta\,\right) =& \psi_{12}^\mathrm{S}\left(\vec\theta\,\right) \Delta_\mathrm{S}\left(m_{12}^2\right) \\
+& \psi_{12}^\mathrm{P}\left(\vec\theta\,\right) \Delta_\mathrm{P}\left(m_{12}^2\right)\\
+& \psi_{13}^\mathrm{S}\left(\vec\theta\,\right) \Delta_\mathrm{S}\left(m_{13}^2\right)\\
+& \psi_{13}^\mathrm{P}\left(\vec\theta\,\right) \Delta_\mathrm{P}\left(m_{13}^2\right)
,\end{eqnarray}
where $\psi_{13}^{\mathrm{S}/\mathrm{P}}$ are the Bose-symmetrized versions of 
$\psi_{12}^{\mathrm{S}/\mathrm{P}}$, with the like-sign pions being 
interchanged.

We find, that $\mathcal{A}_{\mathrm{S} + \mathrm{P}}\left(\vec\theta\,\right)$ 
vanishes everywhere in $\vec\theta$ for a special choice of the dynamic isobar 
amplitudes. Those are described through the following shapes:
\begin{eqnarray}\nonumber
\Delta_\mathrm{S}\left(m_{\xi}^2\right) =& \mathcal{Z}\Big(3m_\xi^2 - m_{\mathrm{K}^-}^2 - 2 m_{\pi^+}^2\\\label{eq:zeroMode}
            &+ \left(m_{\mathrm{D}^+}^2 - m_{\pi^+}^2\right)\frac{m_{\mathrm{K}^-}^2 - m_{\pi^+}^2}{m_\xi^2}\Big)\\\nonumber
\Delta_\mathrm{P}\left(m_{\xi}^2\right) =& \mathcal{Z}
,\end{eqnarray}
with an arbitrary complex-valued coefficient $\mathcal{Z}$. Since  
these two peculiar choices for the dynamic isobar amplitude cancel exactly, 
a shift of the S-wave dynamic amplitude following Eq.~(\ref{eq:zeroMode}) 
can be compensated by a corresponding shift in the P-wave and vice versa. 
Their exact cancellation does not impact the total amplitude 
$\mathcal{A}_{\mathrm{S} + \mathrm{P}}$ anywhere in phase space and a 
PWA fit cannot differentiate these solutions. 
We call a set of dynamic amplitudes for a sub-set of partial waves,
that result in such exact cancellations ``zero mode''.
%A set of functional shapes
%for a subset of dynamic isobar amplitudes that exhibit such exact 
%cancellations is called a ``zero mode''.

Note, that the particular origin of the zero mode in our example 
is caused by the Bose symmetrization and the resulting interference of 
two different combinations of two-particle sub-systems. However, 
such interferences cannot only be caused through Bose symmetrization 
of the final states, but may occur in every process, which allows for 
two or three different two-particle sub-systems to form an isobar. 
If isobars can be formed by a unique two-particle sub-system only, 
the orthogonality of the angular amplitudes ensures the linear independence 
of the fit model. In that case, however, a model-independent PWA is not 
even necessary, since a kinematic binning in $m^2_{\xi}$ suffices to 
extract the resonance content of all partial waves.

Although the zero-mode ambiguity given by the parameter $\mathcal{Z}$ of 
Eq.~(\ref{eq:zeroMode}) can in principle be resolved, as demonstrated in 
Ref.~\cite{zeroModes}, it is advantageous to remove it from the fit model
from the start. This can be achieved through the optimized binning 
described in Sec.~\ref{sec:fitModel}. The optimized binning reflects the 
existence of regions in $m_\xi^2$ with rapidly varying dynamic amplitudes, 
but does not carry information on the amplitudes themselves. 
The best approximation of the shapes in Eq.~(\ref{eq:zeroMode}) by the basis
functions in discrete intervals of $m_\xi^2$ results in partial cancellations
that do not suffice to cause ambiguities in the fit. In particular, the mass 
binning for the various isobaric 
waves will be different. Still, the dynamic binning in $m_\xi^2$ allows for 
narrow resonance structures within an isobar amplitude to be described 
sufficiently well at a price of affordable information loss in regions of 
little variations. With the zero mode now being suppressed, we can reproduce 
the full dynamic amplitudes in the pseudo data containing ground-state and 
excited resonance for the S-wave and the P-waves simultaneously, as shown 
in Fig.~\ref{fig:simul} for the P-wave exemplarily, avoiding leakage effects 
and zero-mode ambiguities.

\section{Summary \& conclusions}
We have demonstrated, that model-independent partial wave analysis 
of three-body final states is a valid method to extract any dynamic 
isobar amplitude of intermediate two-particle sub-systems. However, 
several caveats have to  be respected: 

A) cross-talk between various partial waves, may distort the fit 
results if only a sub-set of waves employed are subjected to the 
model-independent extraction. This cross talk causes distortions 
of the dynamic amplitudes extracted from the data itself by waves 
with imperfect fixed dynamic amplitudes for the isobars.
%results involving only a partially model-independent 
%wave set, can be prone to distorted fit results in model-independent 
%waves due to leakage from non-model-independent waves caused by 
%imperfect descriptions in the dynamic amplitudes.

B) As demonstrated in Ref.~\cite{zeroModes}, the presence of more 
than one freed partial wave can cause exact cancellations of their 
amplitudes at every point in phase space. This manifests as continuous 
ambiguities in the fit.

In this paper we have demonstrated, that an improved model-independent 
functions basis, which relies on an optimized binning in $m_{\xi}^2$ 
for its construction, may suppress these ambiguities and at the same 
time allows to resolve complex resonant features in the isobaric 
amplitude underlying the partial waves. The reason for this is that 
the binning is relatively large in regions without rapidly changing 
resonance content and therefore the model-independent amplitudes are 
unable to approximate the zero mode well enough to cause ambiguities.

Although we exemplarily demonstrated the proposed method for the decay 
$\mathrm{D}^+\to\mathrm{K}^- + \pi^++\pi^+$, we stress that it is 
applicable to a great number of partial-wave analyses of three-body 
final states.

%%%%%%%%%%%%
\tiny{

}
\end{document}